\documentclass[aps,prb,twocolumn,showpacs,preprintnumbers,amsmath,amssymb,superscriptaddress]{revtex4-1}

\usepackage{graphicx}
\usepackage{dcolumn}
\usepackage{bm}
\newcommand{\nc}{\newcommand}
\nc{\be}{\begin{equation}}
\nc{\ee}{\end{equation}}
\nc{\bea}{\begin{eqnarray}}
\nc{\eea}{\end{eqnarray}}
\nc{\bean}{\begin{eqnarray*}}
\nc{\eean}{\end{eqnarray*}}
\nc{\mb}{\mbox}
\nc{\rnc}{\renewcommand}
\nc{\vk}{\mb{\bf k}}
\nc{\vp}{\mb{\bf p}}
\nc{\vn}{\mb{\bf n}}
\nc{\vq}{\mb{\bf q}}
\nc{\rr}{\mb{\bf r}}
\nc{\vz}{\hat {\mb{\bf z}}}
\nc{\vj}{\mb{\boldmath$j$}}
\nc{\vg}{\mb{\boldmath$g$}}
\nc{\x}{\mb{\boldmath$x$}}
\nc{\A}{\mb{\boldmath$A$}}
\nc{\va}{\mb{\boldmath$a$}}
\nc{\vs}{\mb{\boldmath$\sigma$}}
\nc{\vpi}{\mb{\boldmath$\pi$}}
\nc{\nab}{\nabla}
\nc{\X}{\sf x}

\begin{document}

\title{Charge and Spin Reconstruction in Quantum Hall Strips}

\author{Yafis Barlas}
\affiliation{National High Magnetic Field Laboratory and Department of Physics, Florida State
University, FL 32306, USA}

\author{Y. Joglekar}
\affiliation{Department of Physics, Indiana University Prudue University Indianapolis (IUPUI),
Indianapolis, IN 46202, USA}

\author{Kun Yang}
\affiliation{National High Magnetic Field Laboratory and Department of Physics, Florida State
University, FL 32306, USA}

\begin{abstract}
We study the effect of electron-electron interactions on the charge and spin structures of a Quantum Hall strip in a triangularly confined potential. We find that the strip undergoes a spin-unpolarized to spin-polarized transition as a function of magnetic field perpendicular to the strip. For sharp confinements the spin-polarization transition is spontaneous and first develops at the softer side of the triangular potential which shows up as an "eye-structure" in the electron dispersion. For sufficiently weak confinements this spin-polarization transition is preceded by a charge reconstruction of a single spin species, which creates a spin-polarized strip of electrons with a width of the order of the magnetic length detached from the rest of the system. Relevance of our findings to the recent momentum resolved tunneling experiments is also discussed.
\end{abstract}

\pacs{}

\maketitle


\section{Introduction}

Interacting electrons confined to one dimensions exhibit collective spin and charge excitations which lead to exotic phenomena such as spin charge separation~\cite{scsep} and the presence of fractional charges~\cite{frach}. One such example appears at the edges of Quantum Hall (QH) systems in which charged excitations in the bulk are gapped but gapless excitations exist at the edges~\cite{integerandfrac}. The nature of the edge states is sensitive to the details of the edge confinement and the electron-electron interactions. This interplay can lead to rich structures at the edges. In particular it was shown by Chamon and Wen~\cite{chamonwen} that for strong confining potentials the edge of an integer QH state is described by a $Z_{F} =1 $ chiral Fermi liquid, even in the presence of electron-electron interactions. However as the edge confining potential is smoothed beyond certain point the edge  undergoes a reconstruction transition as a strip of electrons move a distance of order 2 magnetic lengths away from the bulk quantum Hall liquid. This leads to the formation of an {\em additional} pair of counter-propagating edge modes, thereby destroying the chiral Fermi liquid nature of charge carriers in favor of the Luttinger liquid. Such a "charge reconstruction" is also expected at the edges of fractional QH states even for sharp confining potentials,~\cite{fracrecon} which has been argued\cite{fracrecon,yangprl03} to be responsible for the lack of universality of electron tunneling exponents at fractional QH edges~\cite{grayson}. Thus far, however, this type of edge reconstruction has not yet been directly observed in experiments.\\
\indent
In an earlier work Dempsey {\it et.al.}~\cite{spinreconhalperin} had  pointed out that edge states can also undergo a second order spin-unpolarized to spin-polarized transition as the confinement potential smoothens. This prediction has been extended to narrow width quantum strips in a perpendicular magnetic field with a parabolic confinement~\cite{spinreconpara}. Recent momentum resolved tunneling experiments~\cite{barak} for parallel wires in a perpendicular magnetic field observe this type of spin-edge reconstruction. Motivated by these experiments we analyze the spin and charge structures of QH strips in a triangularly confined geometry. Furthermore a more interesting question about the interplay between, and possible coexistence of spin and charge reconstruction is addressed within the framework of our model.\\
\indent
Our main results are summarized as follows: as the magnetic field is increased i) a strip of spin polarized electrons develops in the cross section of the strip for sharp confinements, ii) for weak confinements~\cite{footnote} this transtion is preceeded by a charge reconstruction of a single species leading to a detached spin-polarized strip. We use the Hartree-Fock Approximation (HFA) to study the effect of electron-electron interactions on the ground state of electrons in a triangular confinement potential. Hereafter the hard wall of the triangular confinement is referred to as the "hard edge" and linear confinement as the "soft edge". We find that for a fixed confinement slope $\alpha$ the system undergoes a phase transition from a spin-unpolarized to spin-polarized edge as a function of the perpendicular magnetic field with the transition taking place at the soft edge first. For sharp confinements this partial spin polarization of the strip is evident in the dispersion as an "eye-structure", with the spatial separation of the eye increasing as a function of the increasing magnetic field. This spatial separation is proportional to the magnetization of the strip and defines a local order parameter, and our numerical calculations indicate the transition is of second order similar to spin-edge reconstruction in Ref [\onlinecite{spinreconhalperin}]. As the magnetic field is increased the spatial separation of the eye increases and above another critical value of the magnetic field the strip becomes completely spin polarized.
For sufficiently weak confinement the polarization is preceded by a charge reconstruction of single spin species. This reconstruction leads to a formation of a spin polarized strip separated from the rest of the electrons in the strip. The width of this strip increases for increasing magnetic fields and eventually merges with the rest of the electrons in the strip  developing into an "eye-structure". At this point the dispersion is practically indistinguishable form the case of sharp confinement and follows the same sequence; the transition happens at the soft edge first leading to a partially polarized strip eventually leading to a fully spin polarized strip.  For these cases the transition appears to be of first order and is different from the spin-edge reconstruction predicted in Ref [\onlinecite{spinreconhalperin}].
\section{Model}
Earlier studies of edge reconstructions have focused on the structures of isolated edges~\cite{chamonwen,spinreconhalperin}. This is a good approximation for two dimensional QH systems where the edges are separated by the width of the two dimensional structure which is several orders of magnitude larger than the magnetic length. However in the triangular confinement geometry of Ref [\onlinecite{barak}] where the width of the occupied region is of the order several magnetic lengths, the wavefunctions see the effect of both the hard wall and the soft edge. Furthermore the wavefunctions in this confined geometry vanish at the hard wall, which is hard to emulate with Landau level wavefunctions. To account for these important differences we first solve for the eigenfunctions for an electron in a triangular confinement; these eigenfunctions form a basis set to treat electron-electron interactions within the HFA. \\
\indent
We begin by solving for the single particle states for a long quantum strip along the y-direction, with a triangular confinement potential in the x-direction in the presence of a uniform magnetic field ${\bf B}$ in the z-direction. Working in the Landau gauge ${\bf A} = (0,Bx,0) $, taking the confinement potential $V_{c} = \alpha x$ for $x > 0$ and assuming translational invariance in the y-direction $\psi(x,y) = \phi_{\nu}(x) e^{ik_{y} y}$ where $k_{y} =2\pi n/L_{y}$, gives the one-dimensional eigenvalue equation for $ \phi_{\nu}(x)$,
\begin{equation}
\frac{1}{2m} [-\hbar^2 \partial_{x}^{2} + (\hbar k_{y} + \frac{e}{c}Bx)^{2} + 2m \alpha x ]
\phi_{\nu}(x) = \epsilon_{\nu} \phi_{\nu}(x),
\end{equation}
subject to the boundary conditions $\phi_{\nu}(x=0) =0$ and $\phi_{\nu}(x \to -\infty ) =0$. Completing the square and performing a change of variables the equation takes the more familiar form of Weber's differential equation,
\begin{equation}
\label{weber}
[\frac{d^{2}}{d \xi^{2}} - \frac{\xi^{2}}{4} + a] \phi(\xi) = 0,
\end{equation}
where $\xi = \sqrt{2}(x-X)/l_{B}$ is a dimensionless variable written in terms of a generalized guiding center  $X = -k_{y}l_{B}^{2} - \frac{\alpha}{\hbar \omega} l_{B}^{2} $, $l_{B} = \sqrt{\hbar c/(eB)}$ is the magnetic length, and we have defined $ a = 1/(\hbar \omega) (\alpha k_{y} l^{2}_{B} + \alpha^{2}/(2 m\omega^{2}) + \epsilon)$. The original boundary conditions are modified to $\phi_{\nu}(\xi_{0}) =0$, with $\xi_{0} = -\sqrt{2}X/l_{B}$, and $\phi_{\nu}(\xi \to -\infty ) =0$. Eq. (\ref{weber}) allows two independent solutions that can be expressed in terms parabolic cylindrical functions $D_{\nu}(\xi) $ and $D_{-\nu-1}(i\xi)$ where $\nu = a-1/2$. One of them,  $D_{-\nu-1}(i\xi) \sim e^{\xi^2}$ diverges as $\xi \to \pm \infty$ and does not satisfy the boundary conditions. The hard wall boundary condition then reduces to $D_{\nu}(\xi_{0}) = 0 $; solution of this equation gives the order of the parabolic cylinder function $\nu_{q}(\xi_{0})$ for the $q^{th}$ mode of the strip. 
The eigenenergy for the $q^{th}$ mode of the strip can then be expressed as
\begin{equation}
\label{energies}
\epsilon_{q}(k_{y}) = \big( \nu_{q}(\xi_{0}) + \frac{1}{2} \big) \hbar \omega - \alpha k_{y}l_{B}^{2} -\frac{\alpha^2}{2 m \omega^{2}}.
\end{equation}
To be concrete let us analyze the energy of the lowest mode of the strip (i.e. $q=1$). For $\xi_{0} >0 $ one approaches the hard wall giving $\nu_{1}(\xi_{0}) > 1$ which corresponds to a steep increase in the energy dispersion. Alternately for $\xi_{0} <0 $ one starts moving away from the hard wall and $\nu_{1}(\xi_{0}) < 1 $, as we move further still $ \xi_{0} < - 5 $ where the wavefunction can be approximated by the zeroth order Hermite polynomial or the Gaussian (i.e $\nu_{1} =0 $ ). At this point the second term in (\ref{energies}) gives a linear increase in the energy of the Gaussian wavepacket which depends on the confinement potential
$\alpha$. The normalized eigenfunction of the states in the quantum strip,
\begin{equation}
\psi_{\nu_{q}} (x,y) = \frac{2^{1/4}}{\sqrt{L_{y}l_{B} \int_{0}^{\infty}
|D_{\nu_{q}}(u+\xi)|^{2}}}  D_{\nu_{q}} (\frac{\sqrt{2}x}{l_{B}}+ \xi) e^{iky},
\end{equation}
describe the appropriate single particle wavefunction for the $q^{th}$ mode in the strip.
These eigenfunctions form the basis for the interacting problem which will be treated next.\\
\indent
A system of interacting particles of a 2D quantum strip in a uniform magnetic field can be mapped into a one dimensional problem by projecting onto the $q^{th}$ mode of the strip. The Hamiltonian for the interacting theory is
\begin{equation}
\mathcal{H} = \sum_{\lambda, \lambda'} \epsilon_{\lambda, \lambda'} c^{\dagger}_{\lambda}
c_{\lambda'} + \frac{1}{2} \sum_{\lambda_{1}, \lambda_{2},\lambda_{3}, \lambda_{4}}
V_{\lambda_{1}, \lambda_{2},\lambda_{3}, \lambda_{4}}
c^{\dagger}_{\lambda_{1}}c_{\lambda_{2}} c^{\dagger}_{\lambda_{3}}c_{\lambda_{4}}
\end{equation}
where $\lambda_{i} = (\nu_{i},k_{i},\sigma_{i})$ denotes the requisite quantum numbers, $k_{i} = 2 \pi n_{i}/L_{y}$ is the plane wave momentum and is related to the order of the parabolic cylindrical function $\nu_{i}(\xi_{0,i})$ and $\sigma_{i} = \pm$ denotes the spin $\uparrow(\downarrow)$ for the $q^{th}$ mode of the quantum strip. $\epsilon_{\lambda,\lambda'} = \epsilon_{q}(k_{y})\delta_{\sigma,\sigma'} +\sigma \epsilon_{Z}$ where $\epsilon_{Z}$ denotes the Zeeman energy which is taken to be infinitesimal~\cite{footnoteone} in our calculations and
\begin{eqnarray}
\nonumber
V_{\lambda_{1}, \lambda_{2},\lambda_{3}, \lambda_{4}} =
\int d^{2}x_{1} d^{2} x_{2} \psi^{\dagger}_{\lambda_{1}}({\bf x}_{1})\psi_{\lambda_{2}}({\bf x}_{1})\\
\times V(|{\bf x}_{1}-{\bf x}_{2}|) \psi^{\dagger}_{\lambda_{3}}({\bf x}_{2})\psi_{\lambda_{4}}({\bf x}_{2}),
\end{eqnarray}
is the Coulomb interaction projected on the $q^{th}$ mode of the strip, with $ V(|{\bf x}_{1}-{\bf x}_{2}|) =  e^2/(\varepsilon |{\bf x}_{1}-{\bf x}_{2}|)$, where $e$ is the electron charge and $\varepsilon$ is the dielectric constant. Decoupling the four-body interaction within the HFA,  $\langle c^{\dagger}_{k_{1},\sigma} c_{k_{2},\sigma'} \rangle =  \rho^{\sigma} \delta_{k_{1},k_{2}} \delta_{\sigma,\sigma'}$, the HF hamiltonian is $ H_{HF} = \sum_{k,\sigma} \epsilon^{\sigma}_{HF} (k) c^{\dagger}_{k,\sigma} c_{k,\sigma} $ with,
\begin{equation}
\label{energyHF}
\epsilon^{\sigma}_{HF,q}(k) =  \epsilon^{\sigma}_{q}(k) + \sum_{k_{1}}
V_{H}(k,k_{1}) \rho_{q}^{t}(k_{1})- V_{F}(k,k_{1}) \rho_{q}^{\sigma}(k_{1}),
\end{equation}
where $\epsilon^{\sigma}_{k} = \epsilon_{k} + \sigma \epsilon_{Z}$, $\rho^{\sigma}_{q}(k)$ is the wavenumber dependent occupation number for the $q^{th}$ mode and spin state $\sigma$ and $ \rho^{t}_{q}(k) = \rho^{\uparrow}_{q}(k) + \rho^{\downarrow}_{q}(k) $. The Hartree contribution $V_{H}$ is repulsive and captures the long-ranged part of the effective two body interaction, whereas the exchange term $V_{F}$ is attractive and short-ranged. It is the balance of these terms along with the strong confinement already encoded in the non-interacting energy dispersion $\epsilon_{q}(k_{y})$ that determine the self-consistent interacting dispersion. Both the direct and exchange interactions are proportional to $e^{2}/\varepsilon l_{B}$ which defines the natural energy scale of the interacting problem. The Hartree contribution is encoded in matrix element,
\begin{eqnarray}
V_{H}(k_{1},k_{2}) = -\frac{2 e^2}{\epsilon L_{y}} \bigg[
\int^{\infty}_{0} du_{1} \int^{\infty}_{0} du_{2} \log[|\Delta u|] \\
\nonumber
\times |\bar{D}_{\nu_{1}}(u_{1}+\xi_{0,1})|^2 |\bar{D}_{\nu_{2}}(u_{2}+\xi_{0,2})|^2 \bigg],
\end{eqnarray}
where $\Delta u =u_{1}-u_{2}$ and $\bar{D}_{\nu}$ represents the normalized parabolic cylinder function. The logarithmic interaction in $V_{H}$ can be physically understood as the electrostatic repulsion between two line charges which naturally depends on the total density of electrons and is therefore spin-independent. The exchange contribution to the electron self energy is encoded in the exchange matrix element,
\begin{eqnarray}
&& V_{F}(k_{1},k_{2}) = \frac{2 e^2}{\epsilon L_{y}} \bigg[ \int_{0}^{\infty} d u_{1}
\int_{0}^{\infty} du_{2} K_{0}\bigg(\frac{|\Delta \xi||\Delta u|}{2} \bigg) \\ \nonumber
&& \times \bar{D}_{\nu_{1}}(u_{1} + \xi_{0,1}) \bar{D}_{\nu_{2}} (u_{1}+\xi_{0,2})
\bar{D}_{\nu_{1}}(u_{2} + \xi_{0,1}) \bar{D}_{\nu_{2}} (u_{2}+\xi_{0,2}) \bigg],
\end{eqnarray}
where $\Delta \xi = \xi_{0,1}- \xi_{0,2} $ and $K_{0}$ is the  $0^{th}$ order Bessel K function. From the expression it is clear that the exchange matrix element $V_{F}$ depends on the wavefunction overlap, and vanishes for spatially separated orbitals that do not have significant overlap; as a result it is short-ranged.\\

\section{Results and Discussion}
\begin{figure}[t]
\begin{center}
\includegraphics[width=3.0in,height=2.0in]{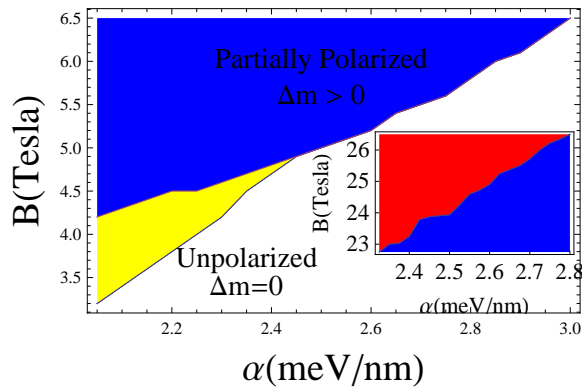}
\caption{Phase diagram of the QH strip. The blue (dark-shaded) region corresponds to a spin-reconstructed phase with partial spin polarization. The yellow (light-shaded) region indicates the coexistence of spin and charge reconstruction. In the inset the blue (dark-shaded) region again corresponds to a partial polarization and the red (light-shaded) region indicates full spin polarization.}
\label{figthree}
\end{center}
\end{figure}

\indent
The Fermi energy of the electrons in the strip depends on the total 1D electron density $\tilde{\rho} =\sum \rho(k_{y}) = N/L_{y}$ where sum is over all occupied states and both spin states and $N$ is the total number of electrons in the strip. We calculate the self-consistent interacting dispersion from (\ref{energyHF}) by adiabatically turning on the interactions. Self-consistency requires that $\rho^{\sigma}_{q}(k) = f(\epsilon^{\sigma}_{HF,q}(k)) $ where $f(\epsilon)$ is the Fermi function. \\
\begin{figure}[t]
\begin{center}
\includegraphics[width=3.5in,height=2.6in]{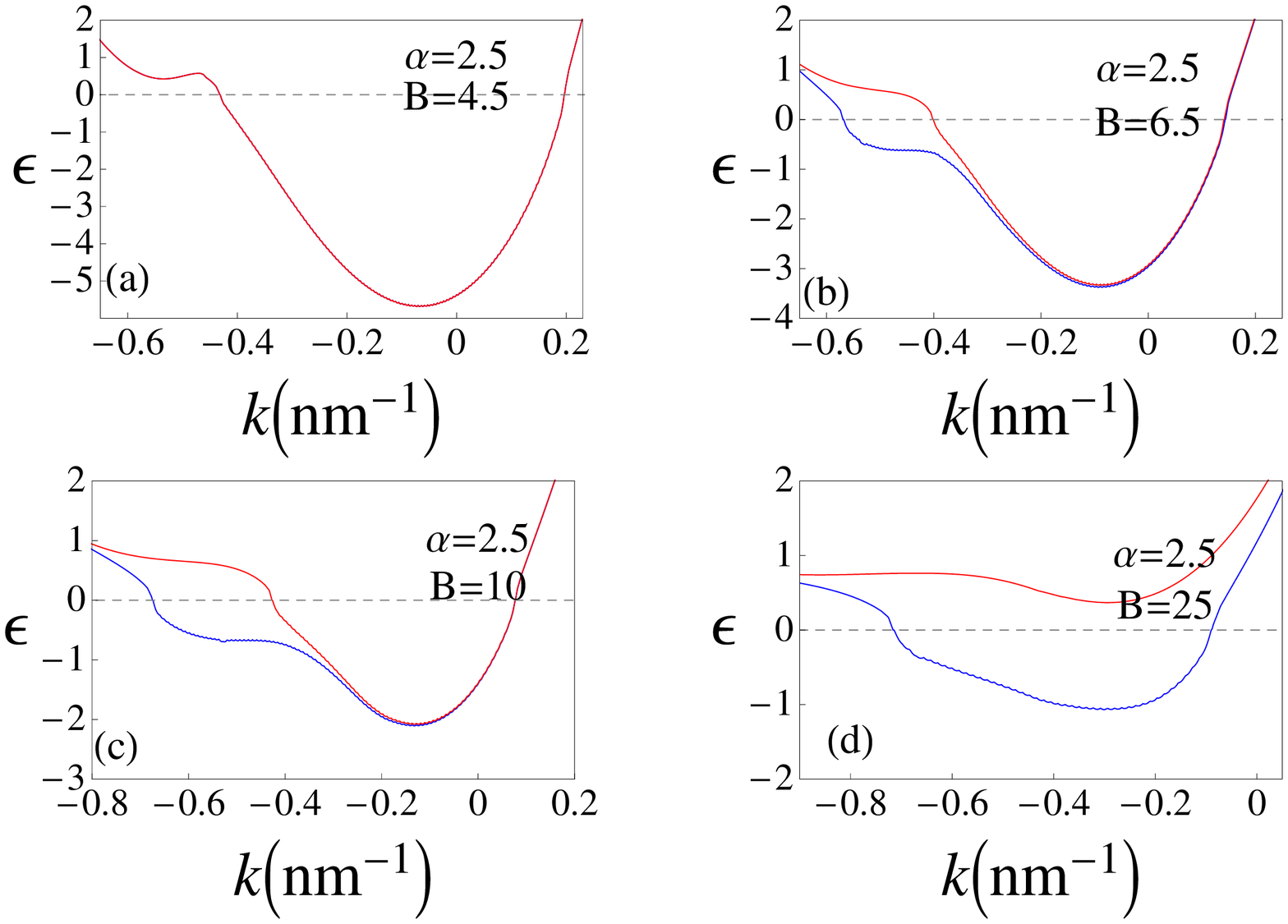}
\caption{Self-consistently calculated spin dispersion for different values of the magnetic field a) 4.5, b) 6.5, c) 10.0, and d) 25.0 T for a fixed confinement potential slope $\alpha = 2.5 meV/nm$ in units of the interaction strength $e^2/(\varepsilon l_{B}) = 4.35 \sqrt{B({\rm Tesla})} meV$. The red(upper) curve corresponds to spin down electron dispersion and the blue (lower) curve corresponds to spin up electron dispersion. The hard wall is placed towards the right and the Fermi energy is normalized to zero and is represented by the dashed horizontal line. The opening of the "eye-structure" due to the magnetization of the strip at the soft side is evident in the dispersion.}
\label{figone}
\end{center}
\end{figure}
\indent
Figures~\ref{figone} a)-d) represent the self-consistently calculated electron dispersion for a range of the perpendicular magnetic fields at a fixed confinement slope $\alpha =2.5 meV/nm$ with a fixed 1D density $\tilde{\rho} = 0.2 nm^{-1}$.~\cite{footnotetwo} The dispersions indicate a transition from a spin-unpolarized to a spin polarized edge with the transition beginning at the "soft edge" of the triangular potential at a critical magnetic field $B_{c}(\alpha)$. The magnetization increases monotonically with the applied magnetic field eventually leading to fully polarized strip. The same sequence is attained when one varies the confinement slope $\alpha$ for a fixed magnetic field $B$, with the transition taking place at a critical value of the confinement slope $\alpha_{c}$. The spin-polarization transition appears as an "eye-structure" in the dispersion where the vertical separation or energy separation of the "eye-structure" is of the order of interaction energy scale in QH systems $ e^2/(\varepsilon l_{B})$. The horizontal separation of the "eye-structure" is proportional to the magnetization and defines a natural order parameter. Our results indicate that the phase transition is continuous.\\
\begin{figure}[t]
\begin{center}
\includegraphics[width=3.5in,height=2.6in]{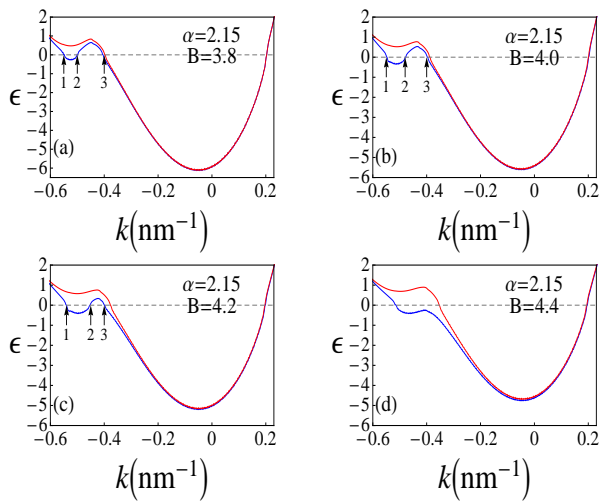}
\caption{Charge reconstruction as a function of the magnetic field a) 3.8, b) 4.0, c) 4.2, and d) 4.4 T for a fixed confinement potential slope $\alpha = 2.15 meV/nm$ in units of the interaction strength $e^2/(\varepsilon l_{B}) = 4.35 \sqrt{B({\rm Tesla})} meV$. The spin polarization is preceded by a single spin species charge reconstruction which leads to a pair of additional crossings $1$ and $2$ indicated by the arrows. As the perpendicular magnetic field is increased crossings $2$ and $3$ annihilate each other leaving a single edge mode. The Fermi energy, hard wall position and electron dispersion labels are the same as in Fig~\ref{figone}.}
\label{figtwo}
\end{center}
\end{figure}
\indent
This type of spin reconstruction for isolated edge states was predicted by Dempsey {\it et.al.}~\cite{spinreconhalperin} for integer QH systems at a total filling factor $\nu = 4$. Even though our geometry is different the physics underlying the spin reconstruction is similar to that of Ref. [\onlinecite{spinreconhalperin}]. The features of the dispersion shown in Figs.~\ref{figone} a)-d) are due to the interplay of confinement and electron-electron interactions. The repulsive Hartree contribution to the interaction forces electrons towards the two edges where the confinement potential opposes this effect. At the hard wall where the confinement is strong the dispersion shows a steep increase in energy whereas for the softer side the energy dispersion varies smoothly. The physics of the spin reconstruction is then associated with the gain in exchange energy the system attains through spin-polarization overcoming the kinetic energy cost. As the magnetic field is increased the electronic dispersion in the strip becomes flatter approaching more closely to a Landau level structure, spin polarization develops when $(\alpha, B)$ crosses the phase boundary of Fig.~\ref{figthree}.\\
\indent
For weaker confinements at the soft side the system exhibits a charge reconstruction for a single spin species~\cite{footnotethree}. This introduces additional modes as evidenced by the single spin branch crossings of the dispersion at Fermi level in Figs~\ref{figtwo} a)-c), and leads to the formation of a {\em detached} spin-polarized strip on the soft-edge of the confinement potential. This type of reconstruction is very different from the spin-type reconstruction discussed earlier. While both lead to a magnetization of the QH strip, the non-monotonic behavior of the dispersion introduces a pair of {\em additional} crossings not seen for the spin-type reconstruction. The transition also becomes first order. As the magnetic field is increased further the second and third crossings move towards each other as shown in Fig.~\ref{figtwo} a)-c) and eventually annihilate each other resulting in the dispersion profile of Fig~\ref{figtwo} d). From this point onwards the dispersions practically become indistinguishable from the spin-type reconstruction scenario. This type of spin-polarization transition is very different from that of Ref [\onlinecite{spinreconhalperin}], and is driven by Chamon and Wen type charge reconstruction~\cite{chamonwen}, which had been predicted for spinless fermion at isolated QH edges. As we see this physics is also present in our geometry for sufficiently weak confinements, although different and richer as it is accompanied by a spin-polarization transition. The full phase diagram for a fixed 1D density $\tilde{\rho} = 0.20 nm^{-1}$ is plotted in Fig \ref{figthree}. The yellow (light-shaded) region indicates the coexistence of spin and charge reconstruction.\\
\\
\section{Relevance to Experiments}
Recently Barak {\it et.al.}~\cite{barak} reported experimental evidence of spin reconstruction of quantum strips in a quantizing perpendicular magnetic field. The electronic dispersion measured by momentum resolved tunneling shows spin-polarization transition at the soft edge of the strip indicated by the opening of an "eye-structure" eventually leading to fully spin polarized strip. Even though the spin reconstruction is observed in the second lowest mode of the strip, qualitative features of the experimental observations are well described by our numerical calculations. Furthermore the presence of the two modes introduces complications due to the orthogonality relation between the first and second mode~\cite{barak} in the tunneling processes. Our calculation points to a much simpler scenario where such a spin reconstruction is possible within the lowest mode of the strip; this avoids the complications associated with the second mode of the strip. More importantly, we predict a new type of spin-charge reconstruction which is driven by charge reconstruction for weaker confinement potentials. This type of reconstruction is more interesting as it leads to {\em additional} edge modes. 
Momentum-resolved tunneling can be used to probe such reconstructions and these additional modes;\cite{akakii04}  the setup of Ref [~\onlinecite{barak}] is particularly well-suited for such purposes.\cite{akakii05}

The authors would like to acknowledge discussions with A. Yacoby and G. Barak and thank them for sharing their data prior to publication. The authors also thank KITP for their hospitality where this work originated. This work was supported in part by NSF grant DMR-1004545 (KY) and the State of Florida (YB).


\begin{thebibliography}{99}
\bibitem{footnote}
This weak confinement has an obvious lower bound as charge reconstruction is not expected in the case $\alpha \leq 0$.
\bibitem{footnoteone}
The Zeeman energy is an order of magnitude smaller than the interaction energy. The spin polarization transition discussed in this paper is spontaneous and purely due to interaction effects, the infinitesimal Zeeman term in our calculations only helps picks a direction in spin space.
\bibitem{footnotetwo}
This choice is based on experimental parameters of Ref~\onlinecite{barak}.
\bibitem{footnotethree}
For higher densities of electrons in the strip $\rho \geq 0.25 nm^{-1}$ we find evidence of charge reconstruction for both spin species. At these densities however the second mode of the strip would be occupied, which is not included in our analysis.
\end{thebibliography}
\end{document}